\documentclass[journal,10pt]{IEEEtran}
\usepackage[T1]{fontenc}
\usepackage{amsmath,amssymb,amsthm,graphicx,booktabs,tabularx,array,cite,url,microtype,enumitem}
\usepackage[hidelinks]{hyperref}
\usepackage{placeins,balance}
\newtheorem{proposition}{Proposition}
\newcommand{\R}{\mathbb R}
\newcommand{\etal}{\textit{et al.}}
\author{Isarar Siddique, Raiyan Siddique, Arshi Niyaz, Jamyang Ugyen Tshomo, and Atish Raza%
\thanks{All authors are with Convolity AI Lab (e-mails: isarar@convolity.com; raiyan@convolity.com; arshi@convolity.com; jamyang@convolity.com; atish@convolity.com). Corresponding author: Isarar Siddique.}}
\title{Evidence Field: Adaptive Covering Programs for Medical Imaging}
\begin{document}
\maketitle
\raggedbottom
\setlength{\abovedisplayskip}{3pt}\setlength{\belowdisplayskip}{3pt}

\begin{abstract}
Medical image classifiers typically return predictions without a native contract for independently reconstructing their spatial decision computation. Evidence Field is a trainable spatial-program architecture that jointly produces class scores, selected covering programs and signed evidence. Its encoder combines absolute and image-relative response fields with associative spatial moments and learned characteristic sums. A typed classifier scores individual regions and parent-conditioned multiplicative groups; a covering recursion selects scale, partition and composition. We establish exact one-fold coverage, optimality within the registered grammar, a bound relating smooth training to hard inference, signed-score conservation and class-centred offset invariance. Direct contractions and their adjoint specify arithmetic and storage scaling on two- and three-dimensional grids. Native output records bind inputs, parameters, supports, types and contextual dependencies to individual and ensemble predictions. In a separately reported histopathology application, patient-separated training and evaluation across held-out centres provide empirical evidence for the architecture. Trained mechanism and corrected-loss comparisons distinguish predictive effects from implementation correctness; independent readers reconstruct descriptors, routes, scores and probability mixtures. The principal contribution is a coupled representation, decision and evidence architecture whose selected computation can be reproduced independently.
\end{abstract}

\section{Architectural objective and contribution}
A class score alone specifies neither spatial evidence nor its reconstruction. Evidence Field (EF) makes both architectural outputs. CNNs aggregate convolutional features~\cite{lecun,resnet}; Transformers aggregate attention-weighted tokens~\cite{vaswani,vit}. EF learns image fields and typed region/group scores, selecting a complete covering program with reconstructible signed contributions.

Absolute/image-relative encoding addresses acquisition dependence~\cite{zech,degrave}, combining regional statistics~\cite{pyradiomics}, radial responses and characteristic sums~\cite{rbf,tancik}. Max-affine types~\cite{maxout}/parent-conditioned products~\cite{tensor} score regions/context. Smooth training and hard deployment share the grammar; independent replay is the principal contribution. Section~\ref{ef:application_evidence} supplies separate application evidence.

\subsection{Relation to spatial and program models}
Spatial pyramids pool fixed cells~\cite{spatialpyramid}; EF selects a class-specific cover. Sum-product networks compose scoped probability factors~\cite{poon}; EF maximizes signed region/group scores. Rule ensembles sum overlapping predicates~\cite{ruleensemble}; EF constrains terminal supports to an exact cover with parent dependencies. Semiring inference precedes exact grammar reduction~\cite{goodman}. EF's coupled encoding/cover/replay contract is distinct from a universal superiority claim.

\section{Complete model and spatial support contract}
\begin{figure*}[!t]
\centering\includegraphics[width=.82\textwidth]{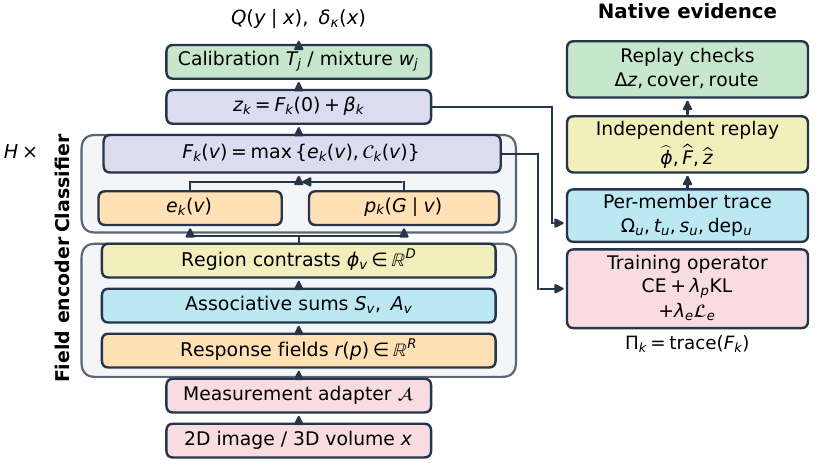}
\caption{EF architecture, read bottom to top. $\mathcal C_k(v)$ is the best partition/packing continuation score; tracing $F_k$ yields program $\Pi_k$. The right branch exposes training and independent replay.}\label{ef:model_architecture}
\end{figure*}

\subsection{Reading Figure~\ref{ef:model_architecture}.}
The adapter converts native samples into six measurement channels $q(p)$. The encoder applies $R$ shared response functions at each location and contracts them over candidate supports. Normalized sums give $f_v$; parent/root contrasts give $\phi_v$. These descriptors feed two score paths: $e_k(v)$ evaluates a single region, while $p_k(G\mid v)$ evaluates a connected group in its parent context. Recursion compares own termination with the best continuation $\mathcal C_k(v)$. The root score plus bias gives $z_k$; calibrated member probabilities form $Q$. The $H\times$ mark denotes grammar depth with shared banks, rather than $H$ independently parameterized classifiers.

The right branch exposes member programs for independent descriptor, route and score reconstruction. Smooth recursion supplies predictions and evidence to the training loss.

\begin{table*}[!t]\centering\small
\caption{Notation used by the architecture and its equations.}\label{ef:notation}
\begin{tabularx}{\linewidth}{@{}lX@{}}\toprule
Symbols & Meaning\\\midrule
$d,N$ & Grid dimension and number of samples in $\Omega$.\\
$R,L,K$ & Response fields, learned frequencies and output classes.\\
$H,A,b,J$ & Grammar depth, partition alternatives, $b=2^d$ children, maximum group order.\\
$M_n,M_g,D$ & Own/group score types and region descriptor width.\\
$v,\pi(v),m_v,\ell_v$ & Region occurrence, parent, midpoint and half-width vector.\\
\bottomrule\end{tabularx}
\end{table*}

Let $x$ occupy a grid $\Omega\subset\mathbb Z^d$, $d\in\{2,3\}$, with $N=|\Omega|$. A declared adapter supplies calibrated measurements and positive axis spacing $s_j$. The scalar contract uses measurement $a$ and a fixed $3^d$ local mean $\bar a$:
\begin{equation}\label{ef:measurements}
 q(p)=[a,\ a-\bar a,\ \partial_z\bar a,\ \partial_y\bar a,\ \partial_x\bar a,
                 (\|\nabla\bar a\|^2+10^{-8})^{1/2}]\in\R^6 .
\end{equation}
Differences are centered internally and one-sided at boundaries; 2D sets the $z$ coordinate/derivative to zero. The optical RGB adapter supplies $q=[o_R,o_G,o_B,|\nabla\bar o_R|,|\nabla\bar o_G|,|\nabla\bar o_B|]$, where $o_c=-\log\operatorname{clip}(x_c,1/256,1)$; channel absorption does not assign stain identity~\cite{vahadane}. Adapters declare units and provenance.

Occurrence $v$ has support $\Omega_v$, mass $\mu_v=|\Omega_v|/N$, parent $\pi(v)$ and depth $h_v\leq H$, $H\geq1$. Each partition creates $b=2^d$ disjoint children. The $A=2$ dictionary registers midpoint and alternating-axis 0.4/0.6 cuts, rounded/clamped inside nondegenerate regions. Fixed occurrences retain parent context; rounding may duplicate alternatives. Scores select partitions/scales with shared descriptor width.

\subsection{Learned tensor banks}
Let $D=3R(10+2L)+3$. Encoder shapes are $(R,6)$ for centres/metrics, $(R)$ for mixtures and $(L,3)$ for frequencies. Own weights have shape $(K,M_n,D)$. Each of $J-1$ interaction-order banks contains affine $(K,M_g,3D-1)$, member/parent $(K,M_g,D-1)$ and scalar $(K,M_g)$ weights. Including bias,
\begin{equation}\label{ef:parameters}
 n_\Theta=13R+3L+KM_nD+(J-1)KM_g(5D-2)+K.
\end{equation}
For $R=8,L=4,K=M_n=M_g=J=2$, $D=435$ and $n_\Theta=10{,}550$. Buffers and 2D-inactive coordinates are separately recorded.
The $13R$ term counts six centres, six metric parameters and one mixture gate per response; $3L$ counts frequencies. Own banks contribute $KM_nD$. Each group type contributes $3D-1$ affine, two $D-1$ projection and one scalar parameter, giving $5D-2$.

\section{Native image-field encoder}
\begin{figure*}[!t]
\centering
\begin{minipage}[t]{.48\linewidth}
\centering\includegraphics[width=.90\linewidth]{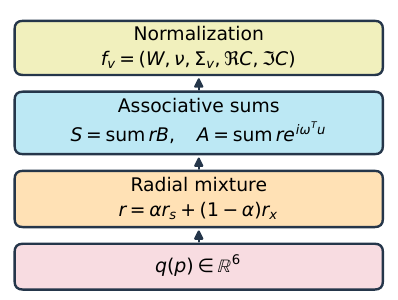}
\caption{Referenced field encoder.}\label{ef:field_panel}
\end{minipage}\hfill
\begin{minipage}[t]{.48\linewidth}
\centering\includegraphics[width=.90\linewidth]{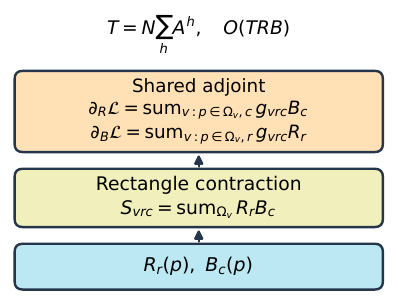}
\caption{Direct contraction and adjoint.}\label{ef:execution_panel}
\end{minipage}
\end{figure*}

\subsection{Absolute and image-relative fields}
Source standardizers define $z_s=\operatorname{clip}((q-\mu_s)/\sigma_s,-6,6)$; current-image moments give
$\tilde q=(q-\mu_x)/\max(\sigma_x,0.03)$ and $z_x=\operatorname{clip}((\tilde q-\mu_{\rm rel})/\sigma_{\rm rel},-6,6)$.
Here $\mu_s,\sigma_s,\mu_{\rm rel},\sigma_{\rm rel}$ are source-fitted buffers; $\mu_x,\sigma_x$ are current-image moments. The variance floor and clipping are part of the reproducible encoder contract.
For learned centre $c_r\in\R^6$, positive diagonal metric
$\Lambda_r=\operatorname{diag}(\operatorname{softplus}\rho_r+10^{-3})$ and mixture $\alpha_r=\operatorname{sigmoid}\eta_r$,
\begin{equation}\label{ef:response_fields}
\begin{aligned}
r(p)={}&\alpha_r e^{-\frac12(z_s-c_r)^T\Lambda_r(z_s-c_r)}\\
&+(1-\alpha_r)e^{-\frac12(z_x-c_r)^T\Lambda_r(z_x-c_r)} .
\end{aligned}
\end{equation}
For positive gain $g$ and offset $t$, $\tilde q(gq+t)=\tilde q(q)$ with inactive variance floor. The input distribution supplies the reference; adapter/absolute branches retain acquisition dependencies~\cite{rbf}.
In Eq.~\eqref{ef:response_fields}, $r(p)$ denotes the $r$th response. Its centre $c_r$ selects a measurement pattern; $\Lambda_r$ controls sensitivity along each channel. The learned $\alpha_r$ blends source-scaled and image-relative distances. These are response coordinates, not predefined anatomical labels.

\subsection{Associative moments and characteristic sums}
Normalized global positions $u\in\R^3$ give
\begin{equation}\label{ef:spatial_sums}
\begin{aligned}
S_{vr}&=\sum_{p\in\Omega_v}r(p)[1,u,\operatorname{vech}(uu^T)]\in\R^{10},\\
W_{vr}&=S_{vr,0},\\
A_{vrl}&=\sum_{p\in\Omega_v}r(p)e^{i\omega_l^Tu(p)},\\
\omega_l&=\omega_{\max}\tanh\kappa_l,\\
C_{vrl}&=e^{-i\omega_l^Tm_v}A_{vrl}/\max(W_{vr},10^{-6}).
\end{aligned}
\end{equation}
Raw sums add over partitions. For $W_\epsilon=\max(W,10^{-6})$, centroid $\nu=S_{1:3}/W_\epsilon$, second-moment matrix $T$ from $S$, and $D_v=\operatorname{diag}(\ell_v)$, define
\begin{equation}\label{ef:regional_features}
\begin{aligned}
f_{vr}=\big[&W/|\Omega_v|,\ D_v^{-1}(\nu-m_v),\\
&\operatorname{vech}(D_v^{-1}(T/W_\epsilon-\nu\nu^T)D_v^{-1}),\\
&\Re C,\Im C\big].
\end{aligned}
\end{equation}
Zero-mass centroid/covariance entries vanish; virtual 2D $z$ half-width is one. Then $f_v\in\R^{R(10+2L)}$, with midpoint-centred phase and $|C|\leq1$. At fixed response and inactive guard,
$\partial C/\partial\omega=i\sum_p r(p)(u(p)-m_v)e^{i\omega^T(u(p)-m_v)}/W$; response gradients include normalization. Finite frequency banks provide spatial sensitivity~\cite{tancik}.
The ten moment coordinates encode mass, three position moments and six unique second moments; $\operatorname{vech}$ stores a symmetric matrix's unique entries. Normalization yields response density, region-relative centroid and covariance. Each complex $C_{vrl}$ contributes two real coordinates sensitive to arrangement at frequency $\omega_l$. The phase shift references the region midpoint.

\subsection{Separating moment-equivalent arrangements}
On five collinear grid positions $u_j=u_0+j\Delta$, let $c_j=(-1)^j\binom4j$. Then $\sum_{j=0}^4c_ju_j^m=0$ for $m=0,1,2,3$, whereas
$\sum_{j=0}^4c_je^{i\omega u_j}=e^{i\omega u_0}(1-e^{i\omega\Delta})^4\ne0$ when $\omega\Delta\notin2\pi\mathbb Z$.
A small $\epsilon c_j$ perturbation of positive response weights preserves mass, centroid and covariance while changing a sampled signature. This separates response-weighted arrangements; diagnostic value requires real-source ablations.

\subsection{Support reuse and referenced features}
Parent/root contrasts yield
\begin{equation}\label{ef:region_contrasts}
\begin{aligned}
\phi_v&=\operatorname{clip}([f_v,f_v-f_{\pi(v)},f_v-f_0,\\
&\hspace{45pt}\mu_v,h_v/H,1],-6,6),\\
D&=3R(10+2L)+3 .
\end{aligned}
\end{equation}
The root references itself. Registry, clipping, guards and coordinate transforms are replayed.

\section{Typed cross-scale group classifier}
\begin{figure*}[!t]
\centering\includegraphics[width=.82\textwidth]{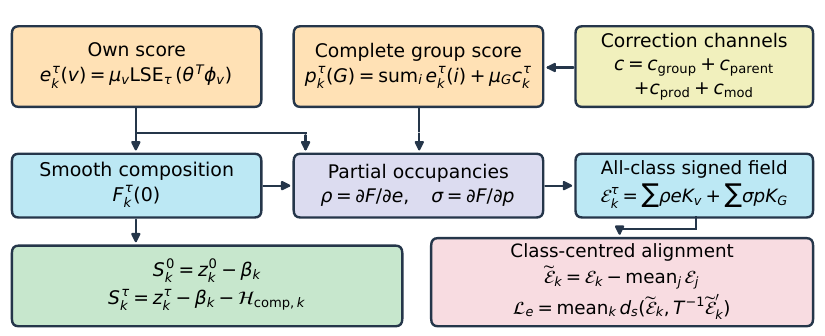}
\caption{Complete-score operands, partial composition occupancies, signed conservation and class-centred alignment. Group scores retain their member anchors.}
\end{figure*}

Class $k$ has $M_n$ own-region types. With $\bar\phi$ denoting the descriptor without its intercept,
\begin{equation}\label{ef:own_scores}
 e_k(v)=\mu_v\max_m\theta_{km}^T\phi_v .
\end{equation}
A production at $v$ permits connected child groups $G$ of orders two to $J$. Connectivity uses face adjacency in the partition's child grid. Let $\mu_G=\sum_{i\in G}\mu_i$,
\[
\begin{aligned}
\phi_G^\mu&=\sum_{i\in G}\mu_i\phi_i/\mu_G,\\
d_G&=\sum_{i\in G}\mu_i|\bar\phi_i-(\phi_G^\mu)_{\rm no\ intercept}|/\mu_G,\\
\psi_{Gv}&=[\phi_G^\mu,d_G,\phi_v].
\end{aligned}
\]
For each order $o=|G|$, $M_g$ types define
\begin{equation}\label{ef:group_scores}
\begin{aligned}
c_{kot}(G\mid v)={}&\xi_{kot}^T\psi_{Gv}\\
&+\gamma_{kot}\bigl[1+\tanh(b_{kot}^T\bar\phi_v)\bigr]\\
&\hspace{15pt}\cdot\prod_{i\in G}\tanh(a_{kot}^T\bar\phi_i),\\
p_k(G\mid v)={}&\sum_{i\in G}e_k(i)+\mu_G\max_t c_{kot}(G\mid v).
\end{aligned}
\end{equation}
Shared projections ensure member-permutation invariance. Order-specific banks and parent affine/product terms supply cross-scale interactions~\cite{tensor}, bound to declared supports and preserved anchors.
Equation~\eqref{ef:own_scores} chooses a class-specific affine type and scales it by region mass. In Eq.~\eqref{ef:group_scores}, $\psi_{Gv}$ combines member mean, disagreement and parent context. The product couples all members; its parent factor modulates that interaction. A complete group score already contains member anchors, so selecting it replaces their recursive subprograms. Signed values may support or oppose a class.

\subsection{Four signed correction channels}
Partition $\xi=[\xi_{\rm group},\xi_{\rm parent}]$ and write $m=\gamma\prod_{i\in G}\tanh(a^T\bar\phi_i)$. The selected correction is exactly $c=\xi_{\rm group}^T[\phi_G^\mu,d_G]+\xi_{\rm parent}^T\phi_v+m+m\tanh(b^T\bar\phi_v)$. Anchors, four corrections and bias replay separately. Parent/root contrasts remain in descriptor ablations.

\subsection{Interaction property}
For fixed type/contrast branch, the product's order-$o$ mixed derivative contains $a^{\otimes o}\mu_G\gamma(1+\tanh b^T\bar\phi_v)$ times member $\operatorname{sech}^2$ factors. Nonzero factors yield interactions absent from independent additive scores.

\section{Adaptive covering recursion and exact execution}
For candidate partition $a$ at $v$, let $\mathcal P_a(v)$ contain all disjoint permitted group packings. Unmatched children are denoted $U(P)$. Leaves use $F_k(v)=e_k(v)$; internal occurrences use
\begin{equation}\label{ef:cover_recursion}
\begin{aligned}
F_k(v)&=\max\{e_k(v),\mathcal C_k(v)\},\\
\mathcal C_k(v)&=\max_{a,P\in\mathcal P_a(v)}
\left[\sum_{G\in P}p_k(G\mid v)+\sum_{u\in U(P)}F_k(u)\right],\\
z_k&=F_k(0)+\beta_k .
\end{aligned}
\end{equation}
Groups terminate member subtrees; unmatched children recurse. Packings are registered before learning; class scores select partition, packing, scale and types.
The continuation in Eq.~\eqref{ef:cover_recursion} is $\mathcal C_k(v)$ in Figure~\ref{ef:model_architecture}. Evaluate from leaves to root: terminate at $v$ or choose a partition/packing and recursively optimal unmatched children. Trace retained winners from the root to recover $\Pi_k$. Covers are class-specific although encoder and grammar are shared.

\begin{proposition}[Cover and registered optimum]
Every selected program covers $\Omega$ exactly once by its score-unit supports. The recursion returns the maximum over this finite grammar.
\end{proposition}
\textit{Proof.} Own supports, disjoint groups and unmatched-child covers partition each region. Depth induction yields one-fold coverage and exhaustive maximization; references add dependencies, not score supports. $\square$

\begin{proposition}[Nested families and rule challenges]
Adding partitions or groups cannot reduce the optimum when common descriptors and scores are fixed. Removing productions cannot increase it. Zero group corrections embed the own-region covering family.
\end{proposition}
\textit{Proof.} The maximized set expands or contracts. With zero corrections, each group equals recursively selectable member anchors. $\square$

For child $i$ in partition $a$, define
\begin{equation}
\begin{aligned}
B_{ka}(i)&=\max\{F_k(i),\max_{G\ni i}p_k(G\mid v)/|G|\},\\
U_k(v)&=\max_a\sum_{i\in{\rm ch}_a(v)}B_{ka}(i).
\end{aligned}
\end{equation}
Member $B$ sums bound unmatched/group scores, including negatives. The own choice skips packing reduction if $e_k(v)>U_k(v)+\epsilon_v$, $\epsilon_v=32\epsilon_{\rm dtype}(1+|e_k(v)|+|U_k(v)|)$, after dense scoring.

\section{Differentiable training, observation and uncertainty}
\subsection{Smooth training and hard decisions}
Cross-entropy learns all declared banks. Type maxima use $\tau\log\sum\exp(s/\tau)$ before mass scaling; composition uses $\tau\mu_v$. Inference uses hard maxima. Source/local validation selects state and calibration.

\begin{proposition}[Soft-to-hard score bound]
For fixed finite descriptors and parameters, depth $H$, at most $A$ alternatives, at most $Q$ packings per partition and $M_n,M_g\geq1$,
\begin{equation}
\begin{aligned}
0&\leq z_k^\tau-z_k^0\leq\varepsilon_\tau,\\
\varepsilon_\tau&=\tau\{\log M_n+\log M_g+H\log(1+AQ)\}.
\end{aligned}
\end{equation}
\end{proposition}
\textit{Proof.} Log-sum-exp adds at most temperature times $\log m$. Type gaps scale by covered mass. Group/unmatched masses partition parent mass; induction adds $\tau\mu_v\log(1+AQ)$ per depth and the type gaps once over terminal mass. Root mass is one. $\square$
A hard class margin above $\varepsilon_\tau$ preserves its winner under this fixed-state smooth reduction.

\subsection{All-class signed score fields}
Let $\operatorname{LSE}_t(s)=t\log\sum_j e^{s_j/t}$ and $\operatorname{LSE}_0(s)=\max_j s_j$. Smoothed own and \emph{complete} group scores $e_k^\tau,p_k^\tau$ enter $\operatorname{LSE}_{\tau\mu_v}$ composition as separate operands, with occupancies
$\rho_{kv}^\tau=\partial F_k^\tau(0)/\partial e_k^\tau(v)$ and $\sigma_{kaG}^\tau=\partial F_k^\tau(0)/\partial p_k^\tau(a,G\mid v)$.
Cloned operands retain parameter ancestry; total derivatives through group anchors would double-count allocation.

Let $U_{aG}=\bigcup_{i\in G}\Omega_i$ and define unit-integral kernels
$K_v(p)=\mathbf1[p\in\Omega_v]/|\Omega_v|$ and
$K_{aG}(p)=\mathbf1[p\in U_{aG}]/|U_{aG}|$.
For every class,
\begin{equation}\label{ef:all_class_signed_field}
 \begin{aligned}
 \mathcal E_k^\tau(p)&=
 \sum_v\rho_{kv}^\tau e_k^\tau(v)K_v(p)
 +\sum_{a,G}\sigma_{kaG}^\tau p_k^\tau(a,G)K_{aG}(p),\\
 S_k^\tau&=\sum_p\mathcal E_k^\tau(p).
 \end{aligned}
\end{equation}
Hard/soft fields allocate selected/expected complete scores, excluding bias; shared references remain dependencies.
The partial occupancies $\rho,\sigma$ measure how much smooth root composition uses each complete operand. Unit-integral kernels spread that operand's signed value across its support. Hard selection makes these occupancies discrete; summing the field recovers the non-bias class score. Context dependencies remain explicit without being counted as extra score supports.

\begin{proposition}[Signed integral and composition entropy]
Let $r_k^\tau(v)$ be the probability of reaching occurrence $v$ and
$\pi_{kv}^\tau$ its local distribution over own termination and all
partition/packing choices. Then
\begin{equation}\label{ef:composition_entropy_identity}
\begin{aligned}
z_k^\tau-\beta_k&=S_k^\tau+\mathcal H_{{\rm comp},k}^\tau,\\
\mathcal H_{{\rm comp},k}^\tau&=
\sum_{v:h_v<H}r_k^\tau(v)\tau\mu_v H(\pi_{kv}^\tau)\geq0 .
\end{aligned}
\end{equation}
At hard inference, $\sum_p\mathcal E_k^0(p)=z_k^0-\beta_k$.
\end{proposition}
\textit{Proof.}
$\operatorname{LSE}_t(s)=\sum_j\pi_js_j+tH(\pi)$ for $\pi=\operatorname{softmax}(s/t)$.
Weight local identities by reach probabilities. Unmatched children recurse; groups/own choices terminate with partial derivative occupancies. Unit-integral kernels allocate scores; hard selection has no entropy residue. $\square$

Type entropy is already contained in smooth unit scores. Only composition entropy is added to their signed integral; both residues vanish under hard reductions. Binary field differences conserve the non-bias class margin.

\subsection{Class-relative signed stability objective}
Let $T$ be an axial quarter turn followed by optional reflection.
Recompute the model on $Tx$ and align its class fields by $T^{-1}$;
the inverse reflects first and rotates inversely.
Define $\widetilde{\mathcal E}_k=\mathcal E_k-K^{-1}\sum_j\mathcal E_j$
pointwise, retaining the common field separately for score accounting.
For equally shaped signed maps $A,B$, use
\begin{equation}\label{ef:shared_signed_l1}
\begin{aligned}
d_{\rm s}(A,B)&=\frac{\|A-B\|_1}{\|A\|_1+\|B\|_1+\epsilon},\quad\epsilon>0,\\
\mathcal L_e&=\frac1K\sum_k
d_{\rm s}\big(\widetilde{\mathcal E}_k^\tau(x),\\
&\hspace{50pt}T^{-1}\widetilde{\mathcal E}_k^\tau(Tx)\big).
\end{aligned}
\end{equation}
Triangle inequality gives $0\leq d_{\rm s}\leq1$; two zero maps give zero.
Centering cancels any aligned class-common field $g$ and commutes with $T^{-1}$. In particular, a common intercept offset adds $C/N$ to fields and $C$ to logits without changing the corrected loss. Its hard integral is the class-centred non-bias logit; original fields retain Eq.~\eqref{ef:composition_entropy_identity}.
With $P=\operatorname{softmax}(z^\tau/T_p)$ and $T_p>0$,
the batch-averaged objective is
\begin{equation}\label{ef:aligned_training_objective}
 \begin{aligned}
 \mathcal L={}&\tfrac12\{\operatorname{CE}(z^\tau(x),y)
                  +\operatorname{CE}(z^\tau(Tx),y)\}
 \\ &+\frac{\lambda_p}{2}
     \{\operatorname{KL}(P(x)\|P(Tx))
        +\operatorname{KL}(P(Tx)\|P(x))\}
 \\ &+\lambda_e\mathcal L_e .
 \end{aligned}
\end{equation}
Setting $\lambda_p=\lambda_e=0$ recovers paired cross-entropy. Differentiating the signed-field objective retains mixed derivatives through score occupancies. Objective settings and their separately trained comparisons belong to Section~\ref{ef:application_evidence}.
CE penalizes label disagreement; symmetric KL aligns the two predictions; $\mathcal L_e$ aligns their class-relative signed fields after undoing $T$. The coefficients control these separate objectives. Centering removes a class-common offset while original fields retain score conservation.

\subsection{Observation and hospital interfaces}
Records bind native inputs/sample counts. Source validation fixes $T_j,\kappa$~\cite{guo,ovadia,selective} for $Q=\sum_jw_j\operatorname{softmax}(z_j/T_j)$, $w_j\geq0$, $\sum_jw_j=1$, $T_j>0$, and review $\delta_\kappa=\mathbf1[\max_kQ_k<\kappa]$. Adaptation declares updated banks, disjoint training/validation patients and state/calibration; coverage/error require measurement.

\section{Native evidence and independent verification}
For declared adapter $\mathcal A$, spacing $s$ and grammar $\mathcal G$,
$\mathcal F_\Theta(x;\mathcal A,s,\mathcal G)=(z,\{\Pi_k,\mathcal E_k^0,\mathcal B_k\}_{k=1}^K)$.
$\Pi_k$ records selected supports/types, $\mathcal E_k^0$ allocates signed scores and $\mathcal B_k$ binds dependencies/input/state. Score supports cover once; contextual dependencies may overlap.

\subsection{Independent reconstruction contract}
Records bind units, supports/types/routes, dependencies, bias and state to
$\widehat z_k=\beta_k+\sum_{u\in\mathcal U_k}s_k(u)$ and $\widehat P_k=\operatorname{softmax}(\widehat z/T)_k$.
Fresh NumPy/SciPy checks descriptors, scores, grammar, one-fold covers, optima and ties. Group anchors count once; references are dependencies. Export all member states/programs, temperatures and weights for $\widehat Q=\sum_jw_j\operatorname{softmax}(\widehat z_j/T_j)$; logit averaging differs.

\subsection{Evaluation binding and evidence challenges}
Input/state and source-lock identities bind predictions. Grammar challenges preserve descriptors; pixel interventions recompute references. Clinical causality requires assessment.

\subsection{Algorithm 1: learning and independent deployment}
Declare $(\mathcal A,s,\mathcal G,\Theta)$; encode $q,r,S,A,\phi$; score/compose $e^\tau,p^\tau$; optimize Eq.~\eqref{ef:aligned_training_objective}. Lock state/calibration; trace $F^0$, export programs/bias and independently replay descriptors, scores, covers, optima and $z,Q,\delta_\kappa$ with binding receipts.

\section{Execution semantics and computational complexity}

\subsection{Direct rectangular contractions and their adjoint}
Direct sums $S_{vrc}=\sum_{p\in\Omega_v}R_r(p)B_c(p)$ use support slices rather than a dense mask. Their shared-array adjoint, for upstream $g_{vrc}$, is
\[
\begin{aligned}
\frac{\partial L}{\partial R_r(p)}
 &=\sum_{v:p\in\Omega_v}\sum_c g_{vrc}B_c(p),\\
\frac{\partial L}{\partial B_c(p)}
 &=\sum_{v:p\in\Omega_v}\sum_r g_{vrc}R_r(p).
\end{aligned}
\]
The adjoint supports mixed derivatives; FP64 fields/sums yield FP32 descriptors. With $B=10+2L$ and $T=\sum_v|\Omega_v|=N\sum_{h=0}^H A^h$ (each partition covers its parent), arithmetic is $O(TRB)$; $T=7N$ at $A=2,H=2$. Field/sum storage is $O(N(R+B)+VRB)$, excluding measurements, scores and autodiff. Dense contraction needs $O(NVRB)$ arithmetic and $O(NV)$ masks.
These arithmetic and storage bounds describe the operator, including its training adjoint.
Figure~\ref{ef:execution_panel} reads forward as response/basis contraction and backward as accumulation onto the same support locations. The restriction $p\in\Omega_v$ is essential: each upstream derivative contributes only where its region used that sample. Shared sums reduce dense-mask work; realized latency still depends on memory access and implementation, as measured separately in Section~\ref{ef:application_evidence}.

\subsection{Score and recursion scaling}
For $V=\sum_{h=0}^H(Ab)^h$ occurrences, $P=A\sum_{h<H}(Ab)^h$ productions and $G$ groups per production, candidate arithmetic is $O(VKM_nD+PGKM_gJD)$ and packing arithmetic is $O(PKQ(b+G))$. At $A=2,H=2,J=2$, the 3D grammar has $(V,P,G,Q)=(273,34,12,108)$; larger order banks remain optional. Fields/classes/supports parallelize; recursion has depth dependencies. At $A=2$, 3D dictionaries grow from $V=273$ to $4369$ and $69905$ at depths 2/3/4. The gate skips packing reductions after scoring, not image encoding; direct storage does not remove registry growth.

\section{Empirical Evidence and Independent Audits}\label{ef:application_evidence}
The mathematical and computational contributions of Sections I--VIII are the subject of this manuscript. The related application manuscript~\cite{ef_hist} reports the shared CAMELYON17 training, classification, calibration and evidence audits. The tables here summarize that existing study; they are not independent replications or an additional clinical validation. The trained mechanism comparisons test architectural choices within the same source-locked protocol.
The separate histopathology companion~\cite{ef_hist} supplies this real-data architecture evidence.

\subsection{Source-only training and evaluation}
Six recipes, seeds 11/29/47: 16 from-scratch epochs (18 fits). CAMELYON17~\cite{camelyon} centres 0/3/4 provide 271,010/24,982/40,004 training/validation/source-test patches; development-disjoint centres 1/2 appear in Table~\ref{ef:companion_results}. All 455,954 patches are included. The wltjr1007 CAMELYON17-WILDS mirror has decoded-image/LFS checks, not authenticated publisher bytes. Source patients split 20/3/3 via retained training IDs/label-blind alternating sorted remaining IDs; hashes are archived. Previously examined targets constitute source-locked replication, not prospective/standard WILDS evaluation.

$96^2$ RGB; EF depth two. AdamW: EF/baseline rates 0.003/0.001, decay $10^{-4}$, cosine schedule, clipping 5, batch 64, unweighted CE. ResNet18/four-block width-128 ViT ($8^2$ patches) are baselines. Source validation selects state/objective/calibration/review. Primary EF: paired CE; signed fits: $T_p=1$, $\lambda_p=0.05$, $\lambda_e=0.02$. Four EF objectives versus one CNN/ViT configuration leave unequal tuning; held-out scores select no model.

\subsection{Measured training resources}
Three-seed 16-epoch A100 80\,GB means: paired-CE EF 1.96\,h, CNN 0.57\,h, ViT 0.48\,h; peak allocations 1.362/0.502/0.363\,GiB. Single-view EF: 1.14\,h, 0.847\,GiB. EF differentiates both views/FP64 contractions; signed losses retain mixed derivatives. Checkpointed/final-successful-attempt logs include concurrent contention, omit uncheckpointed failure overhead; allocations include activations/caches.

\begin{table*}[!t]\centering\small
\caption{Histopathology companion: raw three-member ensemble AUROC / accuracy (\%), fixed argmax. $N$ counts patches, parentheses patients; EF is source-selected paired CE.}\label{ef:companion_results}
\begin{tabular}{@{}lrrrr@{}}\toprule
Evaluation & $N$ (patients) & EF & CNN & ViT\\\midrule
CAMELYON centre 1 & 34,904 (8) & 0.9356 / 84.46 & 0.7352 / 58.94 & 0.8502 / 76.13\\
CAMELYON centre 2 & 85,054 (9) & 0.9453 / 86.77 & 0.7656 / 67.60 & 0.7924 / 67.27\\
\bottomrule\end{tabular}\end{table*}
\begingroup\raggedright EF's 95\% patient-bootstrap AUROC intervals (2,000 resamples): $[0.8311,0.9821]$/$[0.9046,0.9630]$. Paired EF--CNN/EF--ViT: $[0.0972,0.4443]$/$[0.0051,0.2523]$ (centre 1), $[0.1078,0.4319]$/$[0.1137,0.2993]$ (centre 2). Patient resampling retains patch multiplicity. These exploratory, multiplicity-unadjusted comparisons have unequal tuning and few patients.\par\endgroup

\subsection{Trained mechanism and corrected-loss comparisons}
Seven mechanism triplets/three centred-loss fits add 24 evaluated fits (16 epochs); one learning-rate fit is source-only. Rows are separately trained, including direct contractions; 72 prediction files/three statistics reports are complete. External scores do not replace primary EF.
\begin{table*}[!t]\centering\small
\caption{Three-member raw-probability ensemble AUROC; registered $n_\Theta$ per member. No-parent masking leaves 7,074 nominal unmasked coordinates.}\label{ef:trained_ablation}
\begin{tabular}{@{}lrrrr@{}}\toprule
Representation / composition / objective & $n_\Theta$ & Source test & Centre 1 & Centre 2\\\midrule
Full EF, paired CE (reference) & 10,550 & 0.9732 & 0.9356 & 0.9453\\
Full EF, direct contractions & 10,550 & 0.9721 & 0.9368 & 0.9447\\
Moment-based Gaussian closure & 10,550 & 0.9778 & 0.9449 & 0.9669\\
No explicit parent correction & 10,550 & 0.9631 & 0.9436 & 0.9400\\
Node-only covering & 1,858 & 0.9666 & 0.9292 & 0.9457\\
Node-only, enlarged capacity & 10,558 & 0.9659 & 0.9326 & 0.9559\\
Full EF, reduced capacity & 7,074 & 0.9620 & 0.9443 & 0.9365\\
Single fixed partition & 10,550 & 0.9622 & 0.9385 & 0.9554\\
Original signed objective (reference) & 10,550 & 0.9608 & 0.9432 & 0.9451\\
Class-centred signed objective & 10,550 & 0.9598 & 0.9420 & 0.9494\\
\bottomrule\end{tabular}\end{table*}
\par\noindent
Gaussian closure uses moments with the same head/backend. Closure/fixed partitions exceed primary EF's hospital point estimates; learned signatures/adaptive partitions lack established predictive benefit here. Direct-full-minus-closure intervals: $[-0.0175,0.0016]$/$[-0.0494,0.0065]$. Parent removal retains contrasts; node-only removes groups; capacity effects are mixed. Centred-minus-original: $-0.0012$/$+0.0043$, intervals $[-0.0064,0.0021]$/$[-0.0074,0.0260]$. Replay, expressivity and offset invariance are distinct from predictive/stability gains.
\par

\subsection{Native outputs and reproducibility}
Probability/review status accompanies member/class supports/types, signed anchors, four corrections and dependencies (Table~\ref{ef:technical_checks}). Comparators lack this native contract; clinical relevance needs assessment. State/input/patient-bound 126 prediction files yield 949 payloads/336 metrics reproducible within $8.25\times10^{-13}$. Centred-model 50\% source retention transfers to 39.81\%/18.14\%, error 0.99\%/0.86\%; this is no safety guarantee. Variant-specific aligned stability/full replay and matched tuning remain unmeasured.

\subsection{Matched CPU deployment measurements}
Resident three-member ensembles use Apple M5 (16\,GiB), PyTorch 2.7/torchvision 0.22, two CPU threads/one interop and batch one. Three counterbalanced serial passes over 340 real $96^2$ patches yield 1,020 warmed timings/family. EF/CNN/ViT prediction medians are 11.07/10.14/4.39\,ms; p95 values are 11.60/10.81/4.46\,ms. Encoding, recursion and mixture are included; startup/disk input are excluded. Cached PNG-to-probability JSON medians are 11.98/10.41/4.62\,ms (17 inputs; 51 calls); whole-process peak RSS is 440/609/413\,MiB, including framework/setup/validation. Additional EF program/gzip export and independent reader medians are 206.77/103.73\,ms (17 inputs, 102 programs, error $\leq3.22\times10^{-7}$); these are separate stages. CPU repeats are identical; GPU class decisions match. ViT's probability residual $7.02\times10^{-4}$ exceeds the initial $2\times10^{-5}$ tolerance (failure/amendment archived). ViT is fastest here; EF's smaller state does not establish a speed advantage.

\begin{table*}[!t]\centering\small
\caption{Hash-bound CAMELYON17 implementation audits: two source loss inputs, both held-out centres for replay. Finite-difference error is relative; others absolute. Loss checks have no optimizer steps.}\label{ef:technical_checks}
\begin{tabular}{lll}\toprule
Check & Cases & Maximum error\\\midrule
Shared-offset loss & 48 checks, 2 inputs & $1.97\!\times\!10^{-13}$\\
Loss finite differences & 24 & $1.08\!\times\!10^{-9}$\\
Three-seed probability & 340 images, both centres & $2.81\!\times\!10^{-7}$\\
Native prediction logits & 3,060 member-view records & $4.16\!\times\!10^{-6}$\\
Panel descriptors / optima & 6,120 class programs & $1.19\!\times\!10^{-7}$ / $6.38\!\times\!10^{-6}$\\\bottomrule
\end{tabular}\end{table*}

\begin{figure*}[!t]
\centering\includegraphics[width=.82\textwidth]{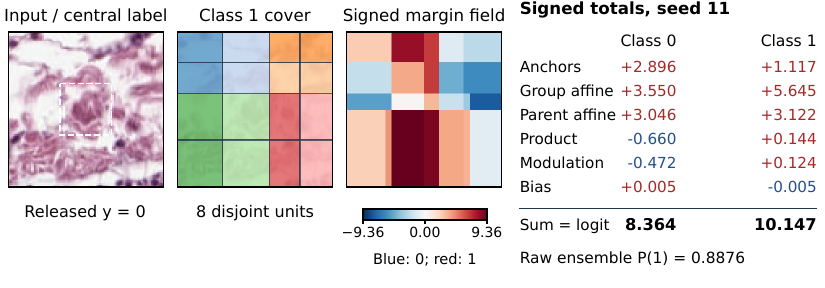}
\caption{Worked evidence: central-$32^2$ $y=0$, seed 11's $96^2$ class-1 cover and signed margin $N(\mathcal E_1^0-\mathcal E_0^0)$. Units are supports, not cells. Raw ensemble $Q_1=0.88756339$ is incorrect.}\label{ef:worked_example}
\end{figure*}
\begingroup\raggedright
\subsection{Public reproduction artifact}
\url{https://github.com/isararsiddique/evidence-field}: code/readers, \href{https://huggingface.co/Isarar/Evidence-Field}{states}, configurations/example. \texttt{download\_artifacts.py} verifies hashes; \texttt{reproduce\_example.py} replays evidence/inference (\texttt{--with-model}). Inventory lists omitted data.\par\endgroup

\subsection{Scope of medical evidence}
Acquisition~\cite{hassan,mackin,berenguer,fortin,vahadane,tellez,ibsi} motivates referencing; lung~\cite{ardila,shen,setio_views,luna,lidc,lungx}, tissue~\cite{camelyon,camelyon16,nuclei,kather_crc,kather_msi,pan_cancer,echle,coudray}, grading~\cite{nagpal,bulten,panda} and supervision/pretraining~\cite{campanella,clam,path_foundation,path_language} require distinct validation. External evaluation~\cite{mckinney,defauw}/bootstrap~\cite{efron} motivate patient-level analyses. The 2D/3D formulation is broader than this 2D application evidence.

\newpage
\subsection{Worked evidence and member accounting}
Figure~\ref{ef:worked_example} binds an unchanged real patch with central-window reference $y=0$ to the source-selected seed-11 model. The class-1 program selects eight disjoint units covering all 9,216 pixels. Its anchors, group affine, parent affine, product and modulation terms are recorded separately from bias. Positive and negative terms describe support for each scored class; their class differences determine the margin. The displayed totals sum to rounded logits 8.364 and 10.147. Full-precision replay has decomposition error at most $5.23\times10^{-7}$ for this member and $2.02\times10^{-6}$ across the three members.

The displayed margin field is rescaled by $N$ for visibility; conservation uses the original unit-integral field and the separate bias difference. Supports explain the executed program, while same-image references remain contextual dependencies. Each ensemble member has its own cover and signed accounting. The reader reconstructs all member logits, applies their recorded temperatures, and mixes probabilities with the declared weights. The raw mixture $Q_1=0.88756339$ is a false positive against the released central-window label. This difficult example shows why a correct numerical explanation is distinct from a correct diagnosis.

\FloatBarrier
\section{Conclusion}
Evidence Field unifies image-field encoding, typed context-aware scores and adaptive covering. Its mathematical contract establishes exact coverage, grammar-relative optimality, soft-to-hard control, signed conservation and class-centred invariance. Native programs expose the computation; direct operators specify arithmetic, storage and adjoints. Empirical audits test these contracts separately from predictive benefit. Application studies evaluate diagnostic tasks while retaining the same architecture.

\section*{Acknowledgment}
Isarar Siddique conceived and planned the study, led the research, and coordinated the project. Raiyan Siddique, Arshi Niyaz, Jamyang Ugyen Tshomo, and Atish Raza assisted with the research. The authors are responsible for the scientific content, verification, and final submission.

OpenAI Codex assisted with drafting and revision (Abstract and Sections I--X), mathematical exposition, figure-layout code (Figs.~1--5), and reproducibility scripts. Input images and reported numerical findings are retained from the bound research records.

The authors declare that they have no conflicts of interest.

\IEEEtriggeratref{34}


\begin{thebibliography}{99}
\bibitem{zech} J. R. Zech, M. A. Badgeley, M. Liu, A. B. Costa, J. J. Titano, and E. K. Oermann, ``Variable generalization performance of a deep learning model to detect pneumonia in chest radiographs: A cross-sectional study,'' \emph{PLOS Medicine}, vol. 15, no. 11, p. e1002683, 2018. doi: \href{https://doi.org/10.1371/journal.pmed.1002683}{10.1371/journal.pmed.1002683}.
\bibitem{degrave} A. J. DeGrave, J. D. Janizek, and S.-I. Lee, ``AI for radiographic COVID-19 detection selects shortcuts over signal,'' \emph{Nature Machine Intelligence}, vol. 3, no. 7, pp. 610--619, 2021. doi: \href{https://doi.org/10.1038/s42256-021-00338-7}{10.1038/s42256-021-00338-7}.
\bibitem{lecun} Y. Lecun, L. Bottou, Y. Bengio, and P. Haffner, ``Gradient-based learning applied to document recognition,'' \emph{Proceedings of the IEEE}, vol. 86, no. 11, pp. 2278--2324, 1998. doi: \href{https://doi.org/10.1109/5.726791}{10.1109/5.726791}.
\bibitem{resnet} K. He, X. Zhang, S. Ren, and J. Sun, ``Deep Residual Learning for Image Recognition,'' \emph{2016 IEEE Conference on Computer Vision and Pattern Recognition (CVPR)}, pp. 770--778, 2016. doi: \href{https://doi.org/10.1109/CVPR.2016.90}{10.1109/CVPR.2016.90}.
\bibitem{vaswani} A. Vaswani \etal, ``Attention is All you Need,'' \emph{Adv. Neural Inf. Process. Syst.}, vol. 30, pp. 5998--6008, 2017. [Online]. Available: \url{https://papers.nips.cc/paper_files/paper/2017/hash/3f5ee243547dee91fbd053c1c4a845aa-Abstract.html}.
\bibitem{vit} A. Dosovitskiy \etal, ``An Image is Worth 16x16 Words: Transformers for Image Recognition at Scale,'' \emph{Proc. Int. Conf. Learn. Represent.}, 2021. [Online]. Available: \url{https://iclr.cc/virtual/2021/poster/3013}.
\bibitem{pyradiomics} J. J. van Griethuysen \etal, ``Computational Radiomics System to Decode the Radiographic Phenotype,'' \emph{Cancer Research}, vol. 77, no. 21, pp. e104--e107, 2017. doi: \href{https://doi.org/10.1158/0008-5472.can-17-0339}{10.1158/0008-5472.can-17-0339}.
\bibitem{rbf} J. Moody and C. J. Darken, ``Fast Learning in Networks of Locally-Tuned Processing Units,'' \emph{Neural Computation}, vol. 1, no. 2, pp. 281--294, 1989. doi: \href{https://doi.org/10.1162/neco.1989.1.2.281}{10.1162/neco.1989.1.2.281}.
\bibitem{tancik} M. Tancik \etal, ``Fourier Features Let Networks Learn High Frequency Functions in Low Dimensional Domains,'' \emph{Adv. Neural Inf. Process. Syst.}, vol. 33, pp. 7537--7547, 2020. [Online]. Available: \url{https://papers.nips.cc/paper_files/paper/2020/hash/55053683268957697aa39fba6f231c68-Abstract.html}.
\bibitem{maxout} I. Goodfellow, D. Warde-Farley, M. Mirza, A. Courville, and Y. Bengio, ``Maxout Networks,'' \emph{Proc. Int. Conf. Mach. Learn. (PMLR)}, vol. 28, pp. 1319--1327, 2013. [Online]. Available: \url{https://proceedings.mlr.press/v28/goodfellow13.html}.
\bibitem{tensor} N. Cohen and A. Shashua, ``Convolutional Rectifier Networks as Generalized Tensor Decompositions,'' \emph{Proc. Int. Conf. Mach. Learn. (PMLR)}, vol. 48, pp. 955--963, 2016. [Online]. Available: \url{https://proceedings.mlr.press/v48/cohenb16.html}.
\bibitem{spatialpyramid} S. Lazebnik, C. Schmid, and J. Ponce, ``Beyond Bags of Features: Spatial Pyramid Matching for Recognizing Natural Scene Categories,'' \emph{CVPR}, pp. 2169--2178, 2006. doi: \href{https://doi.org/10.1109/CVPR.2006.68}{10.1109/CVPR.2006.68}.
\bibitem{poon} H. Poon and P. Domingos, ``Sum-Product Networks: A New Deep Architecture,'' \emph{Proc. Conf. Uncertainty Artif. Intell.}, pp. 337--346, 2011. [Online]. Available: \url{https://www.auai.org/uai2011/accepted.html}.
\bibitem{ruleensemble} J. H. Friedman and B. E. Popescu, ``Predictive learning via rule ensembles,'' \emph{The Annals of Applied Statistics}, vol. 2, no. 3, pp. 916--954, 2008. doi: \href{https://doi.org/10.1214/07-AOAS148}{10.1214/07-AOAS148}.
\bibitem{goodman} J. Goodman, ``Semiring Parsing,'' \emph{Comput. Linguist.}, vol. 25, no. 4, pp. 573--605, 1999. [Online]. Available: \url{https://aclanthology.org/J99-4004/}.
\bibitem{hassan} M. Shafiq-ul-Hassan \etal, ``Intrinsic dependencies of CT radiomic features on voxel size and number of gray levels,'' \emph{Medical Physics}, vol. 44, no. 3, pp. 1050--1062, 2017. doi: \href{https://doi.org/10.1002/mp.12123}{10.1002/mp.12123}.
\bibitem{mackin} D. Mackin \etal, ``Measuring Computed Tomography Scanner Variability of Radiomics Features,'' \emph{Investigative Radiology}, vol. 50, no. 11, pp. 757--765, 2015. doi: \href{https://doi.org/10.1097/rli.0000000000000180}{10.1097/rli.0000000000000180}.
\bibitem{berenguer} R. Berenguer \etal, ``Radiomics of CT Features May Be Nonreproducible and Redundant: Influence of CT Acquisition Parameters,'' \emph{Radiology}, vol. 288, no. 2, pp. 407--415, 2018. doi: \href{https://doi.org/10.1148/radiol.2018172361}{10.1148/radiol.2018172361}.
\bibitem{fortin} J.-P. Fortin \etal, ``Harmonization of cortical thickness measurements across scanners and sites,'' \emph{NeuroImage}, vol. 167, pp. 104--120, 2018. doi: \href{https://doi.org/10.1016/j.neuroimage.2017.11.024}{10.1016/j.neuroimage.2017.11.024}.
\bibitem{vahadane} A. Vahadane \etal, ``Structure-Preserving Color Normalization and Sparse Stain Separation for Histological Images,'' \emph{IEEE Trans. Med. Imaging}, vol. 35, no. 8, pp. 1962--1971, 2016. doi: \href{https://doi.org/10.1109/tmi.2016.2529665}{10.1109/tmi.2016.2529665}.
\bibitem{tellez} D. Tellez \etal, ``Quantifying the effects of data augmentation and stain color normalization in convolutional neural networks for computational pathology,'' \emph{Med. Image Anal.}, vol. 58, Art. no. 101544, 2019. doi: \href{https://doi.org/10.1016/j.media.2019.101544}{10.1016/j.media.2019.101544}.
\bibitem{ibsi} A. Zwanenburg \etal, ``The Image Biomarker Standardization Initiative: Standardized Quantitative Radiomics for High-Throughput Image-based Phenotyping,'' \emph{Radiology}, vol. 295, no. 2, pp. 328--338, 2020. doi: \href{https://doi.org/10.1148/radiol.2020191145}{10.1148/radiol.2020191145}.
\bibitem{ardila} D. Ardila \etal, ``End-to-end lung cancer screening with three-dimensional deep learning on low-dose chest computed tomography,'' \emph{Nature Medicine}, vol. 25, no. 6, pp. 954--961, 2019. doi: \href{https://doi.org/10.1038/s41591-019-0447-x}{10.1038/s41591-019-0447-x}.
\bibitem{shen} W. Shen \etal, ``Multi-crop Convolutional Neural Networks for lung nodule malignancy suspiciousness classification,'' \emph{Pattern Recognition}, vol. 61, pp. 663--673, 2017. doi: \href{https://doi.org/10.1016/j.patcog.2016.05.029}{10.1016/j.patcog.2016.05.029}.
\bibitem{setio_views} A. A. A. Setio \etal, ``Pulmonary Nodule Detection in CT Images: False Positive Reduction Using Multi-View Convolutional Networks,'' \emph{IEEE Trans. Med. Imaging}, vol. 35, no. 5, pp. 1160--1169, 2016. doi: \href{https://doi.org/10.1109/tmi.2016.2536809}{10.1109/tmi.2016.2536809}.
\bibitem{luna} A. A. A. Setio \etal, ``Validation, comparison, and combination of algorithms for automatic detection of pulmonary nodules in computed tomography images: The LUNA16 challenge,'' \emph{Med. Image Anal.}, vol. 42, pp. 1--13, 2017. doi: \href{https://doi.org/10.1016/j.media.2017.06.015}{10.1016/j.media.2017.06.015}.
\bibitem{lidc} S. G. Armato \etal, ``The Lung Image Database Consortium (LIDC) and Image Database Resource Initiative (IDRI): A Completed Reference Database of Lung Nodules on CT Scans,'' \emph{Medical Physics}, vol. 38, no. 2, pp. 915--931, 2011. doi: \href{https://doi.org/10.1118/1.3528204}{10.1118/1.3528204}.
\bibitem{lungx} S. G. Armato III \etal, ``LUNGx Challenge for computerized lung nodule classification,'' \emph{Journal of Medical Imaging}, vol. 3, no. 4, Art. no. 044506, 2016. doi: \href{https://doi.org/10.1117/1.jmi.3.4.044506}{10.1117/1.jmi.3.4.044506}.
\bibitem{camelyon} P. Bandi \etal, ``From Detection of Individual Metastases to Classification of Lymph Node Status at the Patient Level: The CAMELYON17 Challenge,'' \emph{IEEE Trans. Med. Imaging}, vol. 38, no. 2, pp. 550--560, 2019. doi: \href{https://doi.org/10.1109/tmi.2018.2867350}{10.1109/tmi.2018.2867350}.
\bibitem{camelyon16} B. Ehteshami Bejnordi \etal, ``Diagnostic Assessment of Deep Learning Algorithms for Detection of Lymph Node Metastases in Women With Breast Cancer,'' \emph{JAMA}, vol. 318, no. 22, p. 2199, 2017. doi: \href{https://doi.org/10.1001/jama.2017.14585}{10.1001/jama.2017.14585}.
\bibitem{nuclei} K. Sirinukunwattana, S. E. A. Raza, Y.-W. Tsang, D. R. J. Snead, I. A. Cree, and N. M. Rajpoot, ``Locality Sensitive Deep Learning for Detection and Classification of Nuclei in Routine Colon Cancer Histology Images,'' \emph{IEEE Trans. Med. Imaging}, vol. 35, no. 5, pp. 1196--1206, 2016. doi: \href{https://doi.org/10.1109/tmi.2016.2525803}{10.1109/tmi.2016.2525803}.
\bibitem{kather_crc} J. N. Kather \etal, ``Predicting survival from colorectal cancer histology slides using deep learning: A retrospective multicenter study,'' \emph{PLOS Medicine}, vol. 16, no. 1, p. e1002730, 2019. doi: \href{https://doi.org/10.1371/journal.pmed.1002730}{10.1371/journal.pmed.1002730}.
\bibitem{kather_msi} J. N. Kather \etal, ``Deep learning can predict microsatellite instability directly from histology in gastrointestinal cancer,'' \emph{Nature Medicine}, vol. 25, no. 7, pp. 1054--1056, 2019. doi: \href{https://doi.org/10.1038/s41591-019-0462-y}{10.1038/s41591-019-0462-y}.
\bibitem{pan_cancer} J. N. Kather \etal, ``Pan-cancer image-based detection of clinically actionable genetic alterations,'' \emph{Nature Cancer}, vol. 1, no. 8, pp. 789--799, 2020. doi: \href{https://doi.org/10.1038/s43018-020-0087-6}{10.1038/s43018-020-0087-6}.
\bibitem{echle} A. Echle \etal, ``Clinical-Grade Detection of Microsatellite Instability in Colorectal Tumors by Deep Learning,'' \emph{Gastroenterology}, vol. 159, no. 4, pp. 1406--1416.e11, 2020. doi: \href{https://doi.org/10.1053/j.gastro.2020.06.021}{10.1053/j.gastro.2020.06.021}.
\bibitem{coudray} N. Coudray \etal, ``Classification and mutation prediction from non-small cell lung cancer histopathology images using deep learning,'' \emph{Nature Medicine}, vol. 24, no. 10, pp. 1559--1567, 2018. doi: \href{https://doi.org/10.1038/s41591-018-0177-5}{10.1038/s41591-018-0177-5}.
\bibitem{nagpal} K. Nagpal \etal, ``Development and validation of a deep learning algorithm for improving Gleason scoring of prostate cancer,'' \emph{npj Digital Medicine}, vol. 2, no. 1, Art. no. 48, 2019. doi: \href{https://doi.org/10.1038/s41746-019-0112-2}{10.1038/s41746-019-0112-2}.
\bibitem{bulten} W. Bulten \etal, ``Automated deep-learning system for Gleason grading of prostate cancer using biopsies: a diagnostic study,'' \emph{The Lancet Oncology}, vol. 21, no. 2, pp. 233--241, 2020. doi: \href{https://doi.org/10.1016/s1470-2045(19)30739-9}{10.1016/s1470-2045(19)30739-9}.
\bibitem{panda} W. Bulten \etal, ``Artificial intelligence for diagnosis and Gleason grading of prostate cancer: the PANDA challenge,'' \emph{Nature Medicine}, vol. 28, no. 1, pp. 154--163, 2022. doi: \href{https://doi.org/10.1038/s41591-021-01620-2}{10.1038/s41591-021-01620-2}.
\bibitem{campanella} G. Campanella \etal, ``Clinical-grade computational pathology using weakly supervised deep learning on whole slide images,'' \emph{Nature Medicine}, vol. 25, no. 8, pp. 1301--1309, 2019. doi: \href{https://doi.org/10.1038/s41591-019-0508-1}{10.1038/s41591-019-0508-1}.
\bibitem{clam} M. Y. Lu, D. F. K. Williamson, T. Y. Chen, R. J. Chen, M. Barbieri, and F. Mahmood, ``Data-efficient and weakly supervised computational pathology on whole-slide images,'' \emph{Nature Biomedical Engineering}, vol. 5, no. 6, pp. 555--570, 2021. doi: \href{https://doi.org/10.1038/s41551-020-00682-w}{10.1038/s41551-020-00682-w}.
\bibitem{path_foundation} R. J. Chen \etal, ``Towards a general-purpose foundation model for computational pathology,'' \emph{Nature Medicine}, vol. 30, no. 3, pp. 850--862, 2024. doi: \href{https://doi.org/10.1038/s41591-024-02857-3}{10.1038/s41591-024-02857-3}.
\bibitem{path_language} M. Y. Lu \etal, ``A visual-language foundation model for computational pathology,'' \emph{Nature Medicine}, vol. 30, no. 3, pp. 863--874, 2024. doi: \href{https://doi.org/10.1038/s41591-024-02856-4}{10.1038/s41591-024-02856-4}.
\bibitem{mckinney} S. M. McKinney \etal, ``International evaluation of an AI system for breast cancer screening,'' \emph{Nature}, vol. 577, no. 7788, pp. 89--94, 2020. doi: \href{https://doi.org/10.1038/s41586-019-1799-6}{10.1038/s41586-019-1799-6}.
\bibitem{defauw} J. De Fauw \etal, ``Clinically applicable deep learning for diagnosis and referral in retinal disease,'' \emph{Nature Medicine}, vol. 24, no. 9, pp. 1342--1350, 2018. doi: \href{https://doi.org/10.1038/s41591-018-0107-6}{10.1038/s41591-018-0107-6}.
\bibitem{efron} B. Efron, ``Bootstrap Methods: Another Look at the Jackknife,'' \emph{The Annals of Statistics}, vol. 7, no. 1, pp. 1--26, 1979. doi: \href{https://doi.org/10.1214/aos/1176344552}{10.1214/aos/1176344552}.
\bibitem{guo} C. Guo, G. Pleiss, Y. Sun, and K. Q. Weinberger, ``On Calibration of Modern Neural Networks,'' \emph{Proc. Int. Conf. Mach. Learn. (PMLR)}, vol. 70, pp. 1321--1330, 2017. [Online]. Available: \url{https://proceedings.mlr.press/v70/guo17a.html}.
\bibitem{ovadia} Y. Ovadia \etal, ``Can you trust your model's uncertainty? Evaluating predictive uncertainty under dataset shift,'' \emph{Adv. Neural Inf. Process. Syst.}, vol. 32, 2019. [Online]. Available: \url{https://papers.nips.cc/paper_files/paper/2019/hash/8558cb408c1d76621371888657d2eb1d-Abstract.html}.
\bibitem{selective} Y. Geifman and R. El-Yaniv, ``SelectiveNet: A Deep Neural Network with an Integrated Reject Option,'' \emph{Proc. Int. Conf. Mach. Learn. (PMLR)}, vol. 97, pp. 2151--2159, 2019. [Online]. Available: \url{https://proceedings.mlr.press/v97/geifman19a.html}.
\bibitem{ef_hist} I. Siddique, R. Siddique, A. Niyaz, J. U. Tshomo, and A. Raza, ``Evidence Field for Histopathology: Cross-Centre Evaluation and Replayable Spatial Decisions,'' unpublished related manuscript, 2026.
\end{thebibliography}
\end{document}